\documentclass[twocolumn,showpacs,aps,prd,amssymb,nofootinbib,nobibnotes,floatfix,longbibliography]{aastex631}

\usepackage{newtxtext,newtxmath}
\usepackage{ae,aecompl}
\usepackage[mathscr]{euscript}

\usepackage[T1]{fontenc}
\usepackage{hyperref}
\usepackage{amsfonts}
\usepackage{amsmath}
\usepackage{mathrsfs}
\usepackage{epsfig}
\usepackage{graphicx}
\usepackage{url}
\usepackage{hyperref}
\usepackage{float}
\usepackage{color}
\usepackage[utf8]{inputenc} 
\usepackage{epsf}
\usepackage{comment}
\usepackage{slashed}
\usepackage{footmisc}
\usepackage{lipsum}
\definecolor{linkcolor}{rgb}{0.0,0.3,0.5}
\usepackage[caption=false]{subfig}

\usepackage{color}
\definecolor{cerulean}{rgb}{0.0, 0.48, 0.65}

\definecolor{navy}{rgb}{0.2, 0.0, 1.0}

\definecolor{jungle}{rgb}{0.0, 0.5, 0.0}

\usepackage{color}

\definecolor{orange}{rgb}{1,0.5,0}
\definecolor{orangeB}{rgb}{1,0.7,0}

\usepackage[mathlines]{lineno}

\begin{document}

\preprint{APS/123-QED}

\title{Measuring the Transverse Velocity of Strongly Lensed Gravitational Wave Sources  \\ with Ground Based Detectors}

\author{Johan Samsing}
\affiliation{Niels Bohr International Academy, The Niels Bohr Institute, Blegdamsvej 17, DK-2100, Copenhagen, Denmark}

\author{Lorenz Zwick}
\affiliation{Niels Bohr International Academy, The Niels Bohr Institute, Blegdamsvej 17, DK-2100, Copenhagen, Denmark}

\author{Pankaj Saini}
\affiliation{Niels Bohr International Academy, The Niels Bohr Institute, Blegdamsvej 17, DK-2100, Copenhagen, Denmark}

\author{Daniel J. D'Orazio}
\affiliation{Space Telescope Science Institute, 3700 San Martin Drive, Baltimore , MD 21}
\affiliation{Niels Bohr International Academy, The Niels Bohr Institute, Blegdamsvej 17, DK-2100, Copenhagen, Denmark}

\author{Kai Hendriks}
\affiliation{Niels Bohr International Academy, The Niels Bohr Institute, Blegdamsvej 17, DK-2100, Copenhagen, Denmark}

\author{Jose Mar\'ia Ezquiaga}
\affiliation{Niels Bohr International Academy, The Niels Bohr Institute, Blegdamsvej 17, DK-2100, Copenhagen, Denmark}

\author{Rico K.~L.~Lo}
\affiliation{Niels Bohr International Academy, The Niels Bohr Institute, Blegdamsvej 17, DK-2100, Copenhagen, Denmark}

\author{Luka Vujeva}
\affiliation{Niels Bohr International Academy, The Niels Bohr Institute, Blegdamsvej 17, DK-2100, Copenhagen, Denmark}

\author{Georgi D. Radev}
\affiliation{The Niels Bohr Institute, Blegdamsvej 17, DK-2100, Copenhagen, Denmark}

\author{Yan Yu}
\affiliation{The Niels Bohr Institute, Blegdamsvej 17, DK-2100, Copenhagen, Denmark}

\shorttitle{Transverse Velocity of Strongly Lensed Gravitational Wave Sources}
\shortauthors{Samsing et al.}

\date{\today}

\begin{abstract}

Observations of strongly gravitationally lensed gravitational wave (GW) sources provide a unique opportunity for constraining
their transverse motion, which otherwise is exceedingly hard for GW mergers in general.
Strong lensing makes this possible when two or more images of the lensed GW source are observed, as each image essentially allows
the observer to see the GW source from different directional lines-of-sight. If the GW source is moving relative to the lens and observer,
the observed GW signal from one image will therefore generally appear blue- or redshifted compared to GW signal from the other image.
This velocity induced differential Doppler shift gives rise to an observable GW phase shift between the GW signals from the different images,
which provides a rare glimpse into the relative motion of GW sources and their host environment across redshift.
We illustrate that detecting such GW phase shifts is within reach of next-generation ground-based detectors such as
Einstein Telescope, that is expected to detect $\sim$hundreds of lensed GW mergers per year. This opens up completely new ways
of inferering the environment of GW sources, as well as studying cosmological velocity flows across redshift.

\end{abstract}

\section{Introduction}\label{sec:Introduction}

Binary black hole (BBH) mergers have been observed through their emission of gravitational waves (GWs) \citep{2023ApJS..267...29A}, and their
diversity in both
masses \citep{2019ApJ...882L..24A, 2020PhRvL.125j1102A}, spins \citep{2019PhRvD.100b3007Z, 2021PDU....3100791G},
and possibly eccentricity \citep{2019ApJ...883..149A, 2021ApJ...921L..31R, 2022NatAs...6..344G, 2022ApJ...940..171R, 2023arXiv230803822T, 2024arXiv240414286G} indicate that they form in various ways and astrophysical environments. Although several formation channels have been proposed,
including dense stellar clusters \citep{2000ApJ...528L..17P, Lee:2010in,
2010MNRAS.402..371B, 2013MNRAS.435.1358T, 2014MNRAS.440.2714B,
2015PhRvL.115e1101R, 2015ApJ...802L..22R, 2016PhRvD..93h4029R, 2016ApJ...824L...8R,
2016ApJ...824L...8R, 2017MNRAS.464L..36A, 2017MNRAS.469.4665P, Samsing18, 2018MNRAS.tmp.2223S, 2020PhRvD.101l3010S, 2021MNRAS.504..910T, 2022MNRAS.511.1362T}, 
isolated binary stars \citep{2012ApJ...759...52D, 2013ApJ...779...72D, 2015ApJ...806..263D, 2016ApJ...819..108B,
2016Natur.534..512B, 2017ApJ...836...39S, 2017ApJ...845..173M, 2018ApJ...863....7R, 2018ApJ...862L...3S, 2023MNRAS.524..426I},
hierarchical systems \citep{2013ApJ...773..187N, 2014ApJ...785..116L, 2016ApJ...816...65A, 2016MNRAS.456.4219A, 2017ApJ...836...39S, 2018ApJ...864..134R, 2019ApJ...883...23H,
2020ApJ...903...67M, 2021MNRAS.502.2049L, 2022MNRAS.511.1362T},
active galactic nuclei (AGN) discs \citep{2017ApJ...835..165B,  2017MNRAS.464..946S, 2017arXiv170207818M, 2020ApJ...898...25T, 2022Natur.603..237S,
2023arXiv231213281T, Fabj24, 2024ApJ...964...43R},
galactic nuclei (GN) \citep{2009MNRAS.395.2127O, 2015MNRAS.448..754H,
2016ApJ...828...77V, 2016ApJ...831..187A, 2016MNRAS.460.3494S, 2018ApJ...856..140H, 2018ApJ...865....2H,2019ApJ...885..135T, 2019ApJ...883L...7L,2021MNRAS.502.2049L, 2023MNRAS.523.4227A},
very massive stellar mergers \citep{Loeb:2016, Woosley:2016, Janiuk+2017, DOrazioLoeb:2018},
and single-single GW captures of primordial black holes \citep{2016PhRvL.116t1301B, 2016PhRvD..94h4013C,
2016PhRvL.117f1101S, 2016PhRvD..94h3504C}, $\sim 10$ years of GW observations have not yet provided any
clear picture of how each of these channels contribute to the observed merger rate, or if other channels need to be
invoked to decribe the observe properties \citep[e.g.][]{2021ApJ...910..152Z}.

There are reasons to believe that a large
set of observations would help us constrain the underlying
distributions of formation channels, as several key observables are likely to differ, e.g.,
dynamical systems will have distributions of random BH spins \citep[e.g.][]{2000ApJ...541..319K, 2016ApJ...832L...2R,2018ApJ...863...68L} and
a subset of eccentric mergers \citep[e.g.][]{2006ApJ...640..156G, 2014ApJ...784...71S, 2017ApJ...840L..14S, Samsing18a, Samsing2018, Samsing18, 2018ApJ...855..124S, 2018MNRAS.tmp.2223S, 2018PhRvD..98l3005R, 2019ApJ...881...41L,2019ApJ...871...91Z, 2019PhRvD.100d3010S, 2020PhRvD.101l3010S, 2021ApJ...921L..43Z} with distinct distributions in
LIGO-Virgo-KAGRA (LVK) \citep{Samsing18}, DECIGO/TianQin/Taiji \citep[e.g.][]{2017ApJ...842L...2C, 2020PhRvD.101l3010S},
and LISA \citep{2016ApJ...830L..18B, 2018MNRAS.tmp.2223S,2018MNRAS.481.4775D, 2019PhRvD..99f3003K}, whereas binary stars are likely to create BBH mergers with correlated spins on near-circular orbits. However, even constraining sub-channels within the dynamical channel is difficult. For example,
eccentric BBH mergers are expected in both globular clusters (GCs) \citep[e.g.][]{Samsing18},
galactic nuclei (GN) \citep[e.g.][]{2009MNRAS.395.2127O}, active galactic nuclei (AGNs)
disks \citep[e.g.][]{2022Natur.603..237S, Fabj24}, and through Lidov-Kozai oscillations in hierarchical multiple BH stellar systems \citep[e.g.][]{2018ApJ...856..140H, 2019ApJ...881...41L, 2021MNRAS.502.2049L}, although these channels are of completely different nature.

These challenges have led to ideas on how to probe the origin and assembly mechanism of
individual BBH mergers case-by-case \citep[e.g.][]{2014barausse, 2017ApJ...834..200M,  2023MNRAS.521.4645Z, 2024arXiv240305625S},
with the idea of looking for features or imprints in the GW form that can tell something about the nearby environment. 
For example, if the BBH evolves in a gaseous medium, the BHs will experience extra drag-forces, which will show up as a change in the GW
frequency evolution that is different from General Relativity (GR) \citep[e.g.][]{2014barausse,2022garg,2023MNRAS.521.4645Z}.
One can also probe the acceleration of the center-of-mass (COM) of the BBH as it
spirals in \citep[e.g.][]{2011PhRvD..83d4030Y, 2017PhRvD..96f3014I, 2018PhRvD..98f4012R, 2019PhRvD..99b4025C, 2019ApJ...878...75R, 2019MNRAS.488.5665W,
2020PhRvD.101f3002T, 2020PhRvD.101h3031D, 2021PhRvL.126j1105T, 2022PhRvD.105l4048S, 2023PhRvD.107d3009X, 2023arXiv231016799L, 2023ApJ...954..105V, 2024MNRAS.527.8586T}, which can be linked to the surrounding gravitational field, and even the dynamical formation pathway leading to merger \citep{2024arXiv240305625S, 2024arXiv240804603H, 2024arXiv241108572H}. Modulations to the GW waveform could also hint for new physics,
help constraining dark matter properties, or other exotic phenomena.

In this paper we explore how the relative proper (transverse) velocity of BBH mergers can be constrained in strong gravitationally lensed systems,
where two or more images of the GW source are observed. When the wavelength of the GW emission is significantly smaller than the
characteristic size of the lens, gravitational wave lensing and the lensing of light are expected to behave nearly identically in
the geometric optics limit \citep[see e.g.][]{Ezquiaga:2020gdt}. 
Although the first detection of strongly lensed GWs is expected in the coming years~\citep{LIGOScientific:2021izm,LIGOScientific:2023bwz},
the full power of GW lensing will come in the era of Einstein Telescope (ET) and Cosmic Explorer (CE), when 
one expects to see $\sim$ hundreds of strongly lensed events per year \citep[e.g.][]{2022ApJ...929....9X, 2023MNRAS.520..702S}.
This has sparked an enormous recent interest and early searches \citep[e.g.][]{2020arXiv200712709D} with
implications for probing dark matter sub-structures \citep{Tambalo:2022wlm,Caliskan:2023zqm}, wave
optics \citep{2003ApJ...595.1039T, 2023PhRvD.108d3527T, 2024PhRvD.109b4064S},
cosmology \citep{Cremonese:2021puh,Jana:2022shb,Chen:2024xal}, strong fields \citep{2022MNRAS.515.3299G, 2024PhRvD.110d4054P},
and modified gravity \citep{Ezquiaga:2020dao,Goyal:2020bkm,Goyal:2023uvm}.

\begin{figure*}
    \centering
    \includegraphics[width=0.8\textwidth]{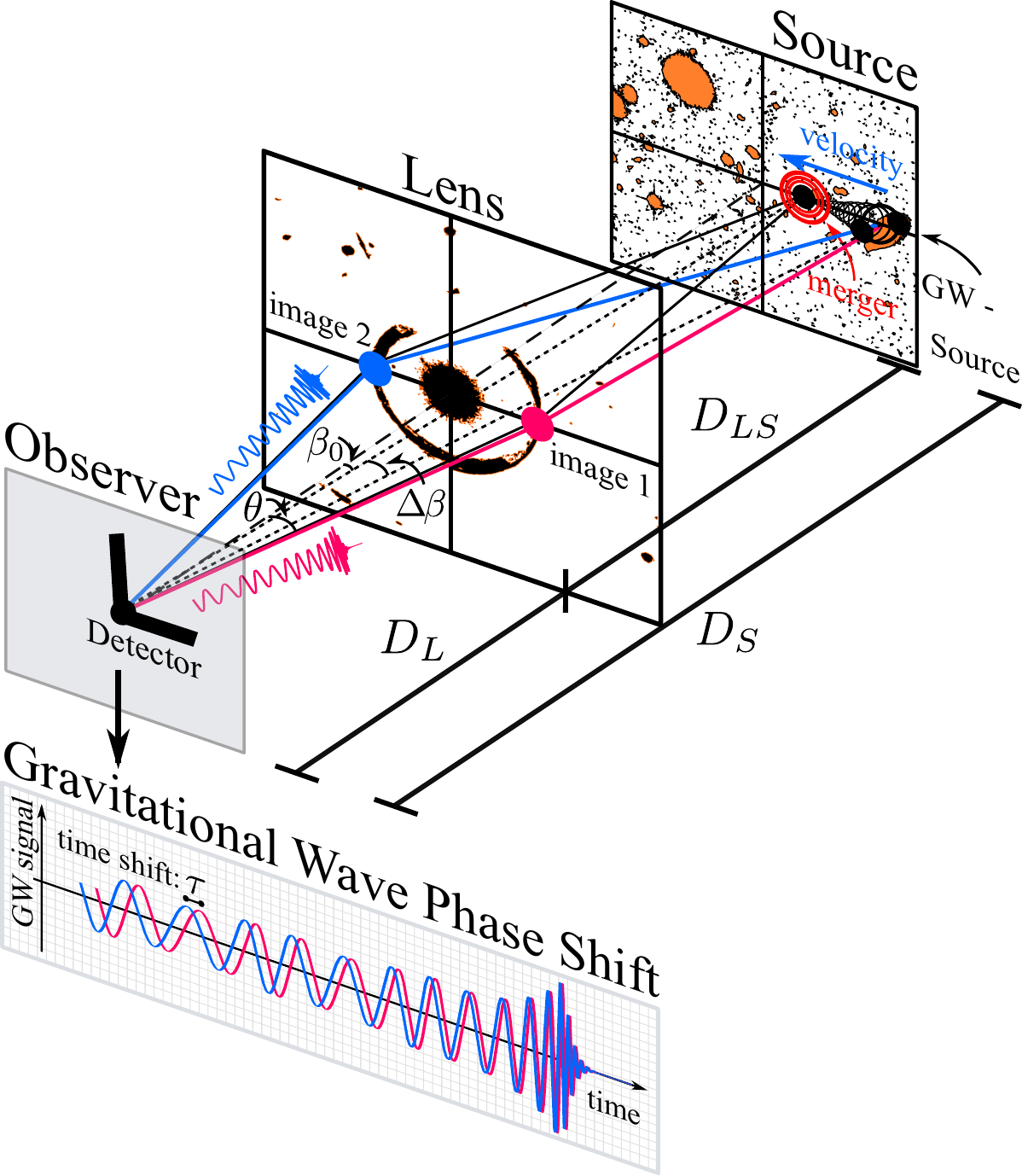}
    \caption{{\bf Illustration of a Strongly Lensed moving Gravitational Wave Source.}
    The {\it Top Panel} of the figure shows our considered Observer-Lens-Source setup, with the detector in the {\it Observer}-plane, the
    gravitational lens in the {\it Lens}-plane, and the lensed GW source in the {\it Source}-plane. In this illustration the GW source
    is moving in the source plane, which takes it from an angle $\beta = \beta_0 + \Delta{\beta}$ to the point of merger at $\beta_0$.
    The transverse velocity vector of the GW source has different projected velocity components along the the lines-of-sight toward the observer, which results in a
    differential Doppler shift between {\it image 1} and {\it image 2}. For GW sources, this will show up as a GW phase shift as illustrated in the {\it Bottom Panel}.
    The detectability for Einstein Telescope is shown in Fig. \ref{fig:GWPH_contour}.}
    \label{fig:ill_lensingBBH}
\end{figure*}

The reason why two or more images of strongly lensed GW source allow for a measure of (transverse) velocity is
that the images essentially define different lines-of-sight towards the BBH from the 
observer (see Fig. \ref{fig:ill_lensingBBH}). This implies that any non-zero velocity of the GW source relative to the
gravitational lens and the observer will project differently along the different lines-of-sight, and will therefore
show up as GW images with slightly different Doppler shifts \citep[e.g.][]{2009PhRvD..80d4009I, 2024PhRvD.109b4064S}.
The relative Doppler shift between pairwise images can be inferred by comparing their GW phase evolution, and the resultant GW phase
shift can therefore be mapped to the relative transverse velocity between the observer, lens and GW source system.

As we here focus on cosmological strongly lensed systems, the relative velocity of the GW source will have components
from its own motion inside its host galaxy, the motion of the galaxy within its environment, e.g. a galaxy cluster, and this
environment relative to the cosmic flow. The contribution from each of these components depend on the redshift,
host galaxy type, and generally the cosmological environment from local to cosmic scales; e.g., a typical galaxy has
an internal velocity dispersion $\sigma \sim 100\ kms^{-1}$, the Milky Way is part of a small local group with
a $\sigma \sim 60\ kms^{-1}$, where galaxies in the large COMA cluster have $\sigma \sim 1000\ kms^{-1}$.
The idea of using strong lensing to probe relative (transverse) motions across the universe is therefore also
a probe of the cosmic flow, how it depends on galaxy types, and how these correlate with the production of
GW sources. We therefore envision this to also be an interesting and independent probe of cosmological dynamics
using GWs in addition to other
methods \citep[e.g.][]{2018ApJ...863L..41F, 2019ApJ...883L..42F, 2021MNRAS.506.2362R,Ezquiaga:2022zkx, 2022ApJ...931...17V, 2022ApJ...932L..19B, 2023MNRAS.522.5546F}. In terms of GW science, the particular interesting question is if the formation of GW
sources are correlated with a particular galaxy or cluster type that can be constrained through their cosmic flow.

With three of more strongly lensed images, the direction of the relative transverse velocity vector can be triangulated, and
we will therefore occasionally refer to this method as {\it Doppler Triangulation}. Variations of this method and moving lens effects
in general have been discussed on several occasions in the literature, both in terms of electromagnetic signals
\citep[][]{1986A&A...166...36K, 1989Natur.341...38C, 1989LNP...330...59B, 2004PhRvD..69f3001W}, as well as for
GW systems \citep[e.g.][]{2009PhRvD..80d4009I, 
2020PhRvD.101h3031D,
2022MNRAS.515.3299G, 2024arXiv241016378Y, 2024PhRvD.109b4064S} with implications 
for deci-hertz detectors such as DECIGO/TianQin/Taiji \citep{2011CQGra..28i4011K, 2016CQGra..33c5010L, 10.1093/nsr/nwx116, 2020PhRvD.101j3027L}.
In this work we quantify and explore the possibilities for the upcoming next-generation ground-based detector ET, which is expected to
see $\sim$hundreds of lensed GW sources per year \citep[e.g.][]{2022ApJ...929....9X}.

The paper is organized as follows. In Sec. \ref{sec:Measuring Transverse Velocity}
derive the basic relations for how a relative velocity between observer, lens, and source relates to a GW phase shift in
the frame of the observer. In Sec. \ref{sec:Results for ET, CE, and LVK} we quantify the measurability of
this GW phase shift for different lensing setups and source properties, through Signal-to-Noise ratio (SNR)
calculations assuming an instrument similar to the proposed ET. 
We conclude in Sec. \ref{sec:Conclusions}. 

\section{Measuring Transverse Velocity}\label{sec:Measuring Transverse Velocity}

Strong lensing of GW sources makes it possible to put constraints on the relative (transverse) velocity of the source,
as each lensed image essentially allow the observer to see the GW source from different
lines-of-sight \citep[][]{2009PhRvD..80d4009I}. If the GW source is moving relative to the lens and the observer, the different
images will show different Doppler shifts, which can be used to triangulate for the velocity vector of
the source. For example, in a static flat Universe, the difference in projected radial velocity, $\Delta{v}$, between two
lensed images separated by a total angle $2\theta$ is to leading order
\begin{equation}
\Delta{v} \approx 2 \theta v,
\end{equation}
where $v$ is the transverse velocity of the GW source relative to the lens and observer (see Fig. \ref{fig:ill_lensingBBH}).
This gives rise to a differential Doppler shift, $\sim \Delta{v}/c$, that can be linked to a displacement in
angular phase of the received GW signals, i.e. a GW phase shift, $\delta{\phi}$, that can be expressed as,
\begin{equation}
\delta{\phi} \approx 2\pi f t \Delta{v}/c \approx 4\pi\theta f t v/c,
\label{eq:deltaphi_simple}
\end{equation}
where $f$ and $t$ is the GW frequency and time, respectively.
The observable GW phase shift between images can therefore be used to infer the relative transverse velocity of the source,
if enough cycles can be build up during the observation through the time $t$ and GW frequency $f$.

In the following we describe the components that go into measuring the relative transverse velocity of the GW source in
an expanding flat Universe assuming two images, with three or more following trivially. Throughout the paper we assume the thin-lens approximation,
the geometric optics approximation, we describe the GW source as a point source, and generally assume a simple lens system
that only leads to smooth changes as the source, lens and observer are moving relative to each other.

\subsection{Relative Velocities in an Expanding Universe}\label{sec:Relative Velocity in an Expanding Universe}

The relative Doppler shift between the images that makes Doppler Triangulation possible, is not just sensitive to the
motion of the source in the source plane, but depends generally on the relative velocity between the observer (O),
lens (L) and source (S) (see Fig. \ref{fig:ill_lensingBBH}). Similar effects will therefore also be seen if the lens moves
and the source is at rest, which is a configuration that often is refereed to as the moving lens effect \citep[e.g.][]{2024arXiv241016378Y}.
This effect has a particular interest for cosmic-microwave-background (CMB) experiments for e.g. probing the
large-scale transverse velocity field \citep[e.g.][]{2021PhRvD.104h3529H, 2024arXiv240816055B}.
Below we derive and illustrate how the observer, lens, and source velocities combine in an expanding universe into an effective velocity.

In our setup we consider three planes denoted the {\it Observer plane}, the {\it Lens plane}, and
the {\it Source plane}, as shown in Fig. \ref{fig:ill_lensingBBH}. In an expanding universe, the relative velocities
translate non-trivially between these planes, as measures of time and length scale with the expansion,
or redshift, $z$. In the following we will project the different velocity contributions from the planes into an effective velocity of the
source in the source plane, relative to the lens and the observer. For this we start by defining the transverse velocities
in each plane as,
\begin{equation}
v_s = \frac{ds_s}{dt_s},\ v_L = \frac{ds_L}{dt_L},\ v_O = \frac{ds_O}{dt_O}.
\end{equation}
where $ds$ and $dt$ denote transverse length- and time-differentials, and the subscripts `$S$', `$L$', and `$O$' refer to the
source, lens and observer planes, respectively.
To translate these velocities into an effective velocity of the source in the source plane, we now consider
the following projections of the length differentials onto the source plane \citep[e.g.][]{1986A&A...166...36K},
\begin{equation}
ds_L' = - \frac{D_S}{D_L}ds_L,\ ds_O' = \frac{D_{LS}}{(1+z_L)D_L}ds_O,
\end{equation}
where the super-script (') denotes that the (unprimed) quantity has been mapped to the source plane,
$D$ is the angular diameter distance,
and $z$ is the redshift. Similarly, time differentials projected onto the source plane are,
\begin{equation}
dt_L' = - \frac{1+z_L}{1+z_S}dt_S,\ dt_O' = \frac{1}{1 + z_S}dt_O.
\label{eq:dtO}
\end{equation}
By combining these differentials, one can now define an effective transverse velocity,
\begin{equation}
v' = v_S - \frac{1+z_S}{1+z_L}\frac{D_S}{D_L}v_L + \frac{1+z_S}{1+z_L}\frac{D_{LS}}{D_{L}}v_O,
\label{eq:v_prime}
\end{equation}
which can be considered as the velocity of the source relative to a setup with lens and observer being fixed.
Below we continue by illustrating how this effective velocity, $v'$, of the source in the source plane can be
turned into an observable GW phase shift.

\begin{figure}
    \centering
    \includegraphics[width=0.5\textwidth]{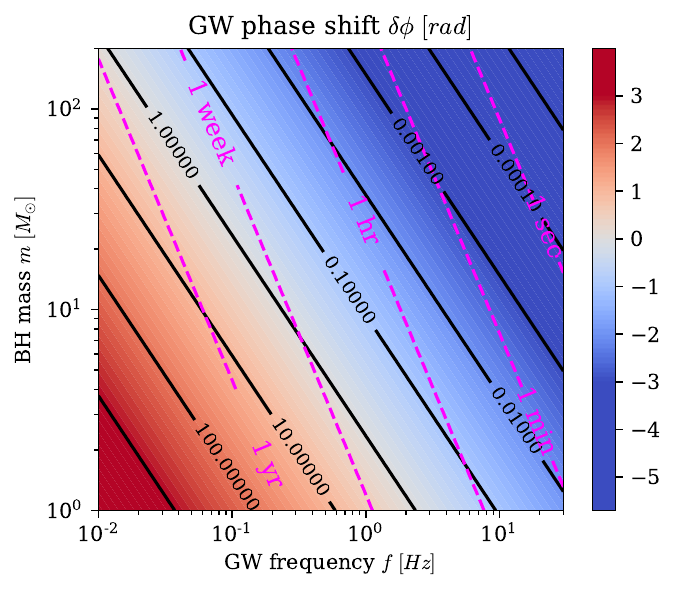}
    \caption{{\bf GW Phase Shift from Strongly Lensed GW sources.} The background {\it colored contours} with color bar to the right,
    show the $\log_{10}$ of the velocity induced GW phase shift, $\log_{10}(\delta{\phi})$, between two strongly lensed images
    derived using Eq. \ref{eq:dphi_general}, as a function of the GW
    frequency $f$, and object mass $m$, all defined in the observer frame. The {\it black contour lines} show the
    value $\delta{\phi}$. The overlayed {\it pink contour lines} show the corresponding GW merger time
    from Eq. \ref{eq:tmerg}. We have assumed $\theta = 25''$, and $v_d = 1000\ kms^{-1}$.
    As seen, for our chosen values, the GW phase shift can accumulate up to $\sim 1.0$ radian for next-generation ground-based
    detectors ($\text{min}(f) \approx 1-5\ Hz$), whereas for LVK one expects closer to $\sim 0.01$ for a typical $\sim 10M_{\odot}+10M_{\odot}$ BBH merger. Corresponding $\delta$SNR results for ET are shown in Fig. \ref{fig:GWPH_contour}.}
    \label{fig:GWPH_contour_fm}
\end{figure}

\subsection{Gravitational Wave Phase Shift}\label{sec:Gravitational Wave Phase Shift}

Consider two GW signals that have been observed in a strong lensing event and subsequently being aligned such that
their time of merger coincides (see Fig. \ref{fig:ill_lensingBBH}). If one signal is Doppler shifted relative to
the other, the two signals will show increasingly larger temporal displacement from each other when going backwards in
time from the point of merger. If we denote this time displacement by $\tau$ at a given GW frequency $f$ defined in the observer frame,
then the GW phase shift can be approximated by,
\begin{equation}
\delta{\phi} = 2\pi{\tau}/{T} = 2\pi f \tau, 
\label{eq:dphi_def}
\end{equation}
where $T$ is the time between GW cycles \citep[e.g.][]{2024arXiv240305625S, 2024arXiv240804603H, 2024arXiv241108572H}.
In this formalism, $\tau$ is the parameter that encodes information about the lensing system and the relative motion of the GW source.
More information is available through the time delay and relative image magnification;
however this is not directly relevant for our GW phase shift estimates considered here.

The time displacement $\tau$, can e.g. be calculated using the time delay formalism in strong lensing systems.
In a strongly lensed system, the extra time it takes for a wave packet to reach the observer along its deflected path relative to
an undeflected is given by \citep[e.g.][]{2002ApJ...568..488O},
\begin{align}
        \Delta{t}(\theta, \beta_0, \Delta{\beta}) & \approx \frac{D_LD_S(1+z_L)}{2cD_{LS}}\left( |(\theta - \beta_{0}) - \Delta{\beta}|^{2}\right), \nonumber\\
	                           & \approx \frac{D_LD_S(1+z_L)}{2cD_{LS}}\left((\theta^{2} - 2{\theta}{\beta_0}) - 2{\theta}\Delta{\beta}\right),
    \label{eq:dt_timedelay}
\end{align}
where $\theta$ denotes the apparent angular position relative to the position of the lens,
$\beta_0$ denotes the angular position of the GW source at merger, and $\beta = \beta_0 + \Delta{\beta}$ denotes the GW source
position at an earlier time. 
This relation implies that if the GW source moves from an angular position $\beta_0 + \Delta{\beta}$ to $\beta_0$, then the change in time delay will be
$\delta{t} \approx \Delta{t}(\theta, \beta_0, \Delta{\beta}) - \Delta{t}(\theta, \beta_0, \Delta{\beta} = 0)$ (see Fig. \ref{fig:ill_lensingBBH}).
This change is not directly possible to measure from a single image of a GW merger, as it will be degenerate with the
velocity, redshift, or mass of the source. However, with two images the change in time delay will result in that
the GW signal in one of the images will arrive earlier or later compared to the other image. This time difference is what we
denote as $\tau$, which then is $\tau \approx 2\delta{t}$. By now using Eq. \ref{eq:dt_timedelay} we find,
\begin{equation}
\tau \approx \frac{2D_LD_S(1+z_L)}{cD_{LS}}{\theta}\Delta{\beta}.
\label{eq:tau_td}
\end{equation}
If the source is moving relative to the lens and observer then,
\begin{equation}
\Delta{\beta} = v't_O'/D_S,
\end{equation}
and together with relations from Eq. \ref{eq:dtO}, Eq. \ref{eq:v_prime}, and Eq. \ref{eq:tau_td},
we can now write Eq. \ref{eq:dphi_def} as,
\begin{equation}
\delta{\phi} = 4 \pi {\theta} f t v_d/c
\label{eq:delta_ftcv}
\end{equation}
where $v_d$ is an `effective Doppler velocity' given by \citep[see e.g.][]{2009PhRvD..80d4009I},
\begin{equation}
v_d = v_O + \frac{D_L}{D_{LS}}\frac{1+z_L}{1+z_S}v_S - \frac{D_S}{D_{LS}}v_{L}.
\label{eq:vd}
\end{equation}
Note here that the form of Eq. \ref{eq:delta_ftcv} is identical to our simple estimate given by Eq. \ref{eq:deltaphi_simple},
but with $v$ given by $v_d$ from above.

These relations provide the GW phase shift as a function of the GW frequency, $f$ and time $t$ in the observer
frame. However, in our setup we assume the GW merger is also observed as this allows for an accurate time alignment of the GW
forms received from the two images, which implies that $f$ is related to $t$ through the
merger time. Using the relations from \cite{Peters64}, we can express the merger time, $t_m$, as a function of
GW frequency, $f$, as,
\begin{equation}
t_m = \frac{2^{4/3} 5}{\pi^{8/3} 512} \frac{c^{5}}{G^{5/3}} \times \frac{1}{m^{5/3}f^{8/3}},
\label{eq:tmerg}
\end{equation}
where we have assumed the case where each of the two merging compact objects have an equal mass $m$.
By now substituting $t$ from Eq. \ref{eq:delta_ftcv} with this expression for $t_m$ one finds,
\begin{align}
	\delta{\phi}   & = \frac{2^{4/3} 5}{\pi^{5/3} 128} \frac{c^{5}}{G^{5/3}} \times \frac{{\theta}}{m^{5/3}f^{5/3}}\frac{v_d}{c}, \nonumber\\
                       & \approx 5\ \text{rad} \left(\frac{\theta}{25''}\right) \left(\frac{M_{\odot}}{m} \right)^{5/3} \left(\frac{Hz}{f}\right)^{5/3} \left(\frac{v_d/kms^{-1}}{1000} \right)
    \label{eq:dphi_general}
\end{align}
where $m$ and $f$ are both defined in the observer frame.
Fig. \ref{fig:GWPH_contour_fm} shows the GW phase
shift $\delta{\phi}$ (Eq. \ref{eq:dphi_general}), and the corresponding merger
time (Eq. \ref{eq:tmerg}), as a function of the GW frequency, $f$ (x-axis) and
BH mass $m$ (y-axis) in the observer frame, for $\theta = 25''$ \citep[e.g.][]{2009MNRAS.392..930O},
and $v_d = 1000\ kms^{-1}$. As seen, given the right combination of $m$, $f$ and detector sensitivity range,
the GW phase shift can be significant with great prospects for measurebility \citep[see also][]{2009PhRvD..80d4009I}, as we quantitfy in
Sec. \ref{sec:Results for ET, CE, and LVK} below.

\section{Detectability}\label{sec:Results for ET, CE, and LVK}

\begin{figure}
    \centering
    \includegraphics[width=0.49\textwidth]{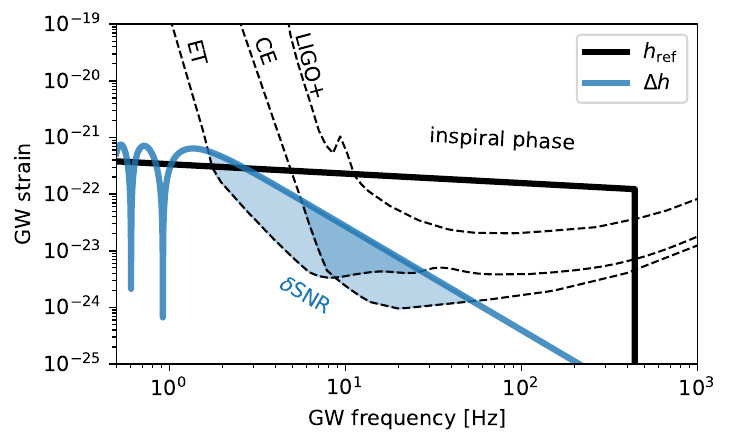}
    \caption{{\bf Strain Evolution and Detector Sensitivity.}
    The {\it solid black} line shows two overlapping GW strains, $h_1$ and $h_2$, that differ by a velocity induced GW phase shift, $\delta{\phi}$, where the
    {\it solid blue} line shows their difference $\Delta{h} = h_1 - h_2$. The specific parameters for this system is
    $z_{L}=0.5$, $z_{\rm S}=1$, $M = 5 M_{\odot}$, $\mu=10$, $v_{S}=3000\ kms^{-1}$, $M_{\rm L}= 3\times 10^{14}\ M_{\odot}$. The sensitivity curves for
    ET, CE and LIGO are shown with {\it dashed lines}. Both the planned ET and CE will operate with a sensitivity to realistically probe this effect for
    $\sim$hundreds of astrophysical lensed GW sources.
    }
    \label{fig:dsnr}
\end{figure}

\begin{figure*}
    \centering
    \includegraphics[width=0.49\textwidth]{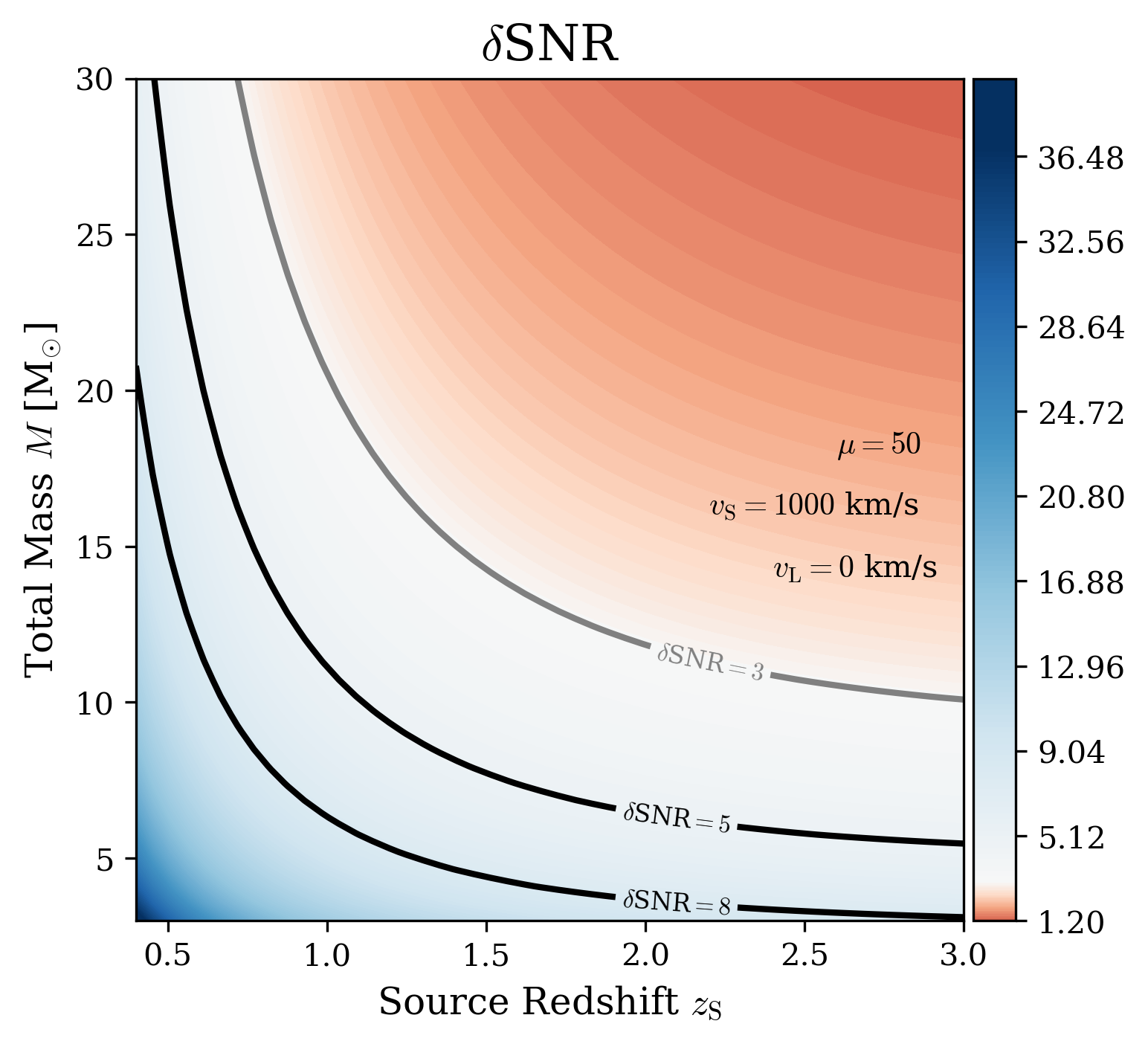}
    \includegraphics[width=0.49\textwidth]{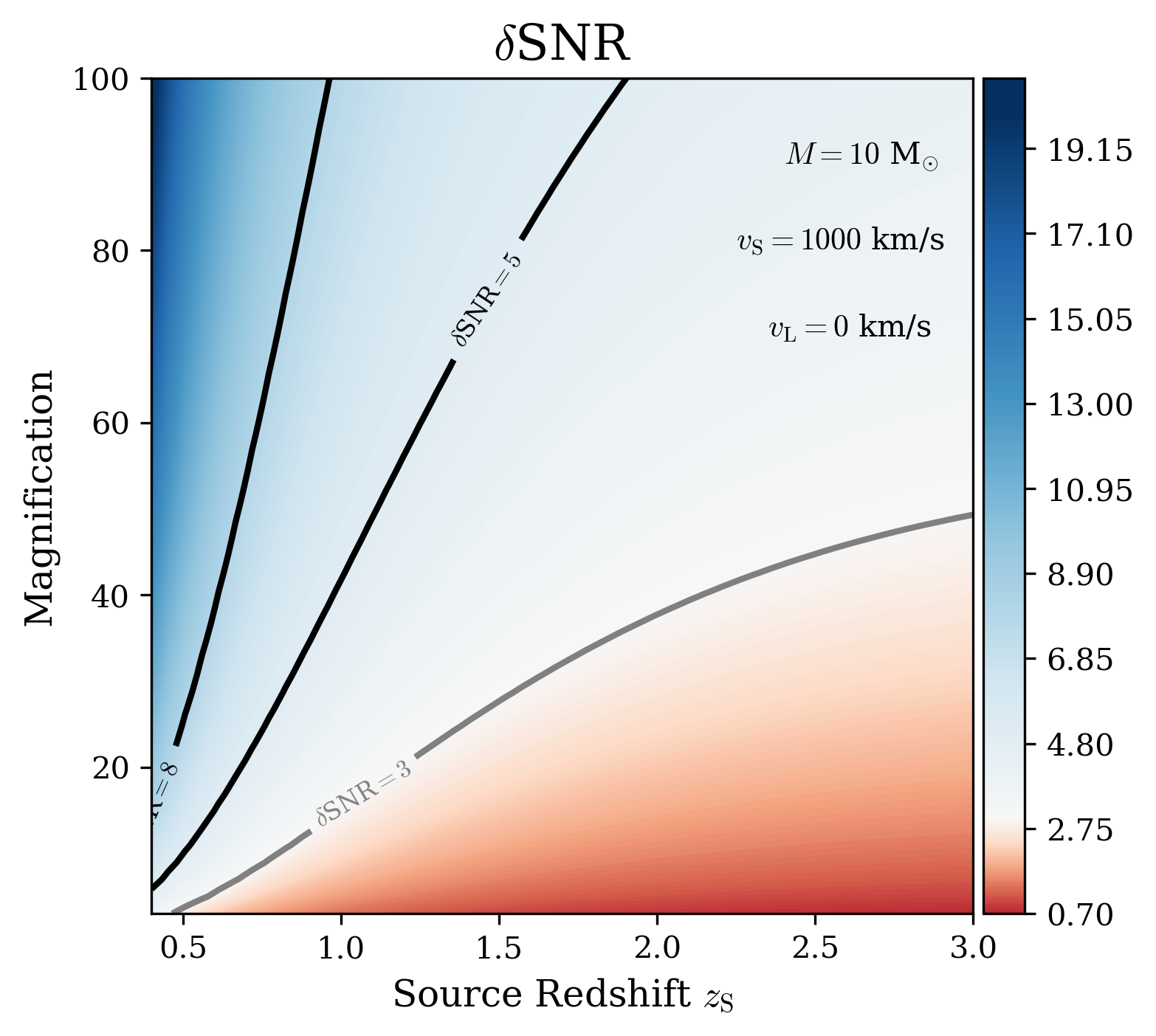}
    \includegraphics[width=0.49\textwidth]{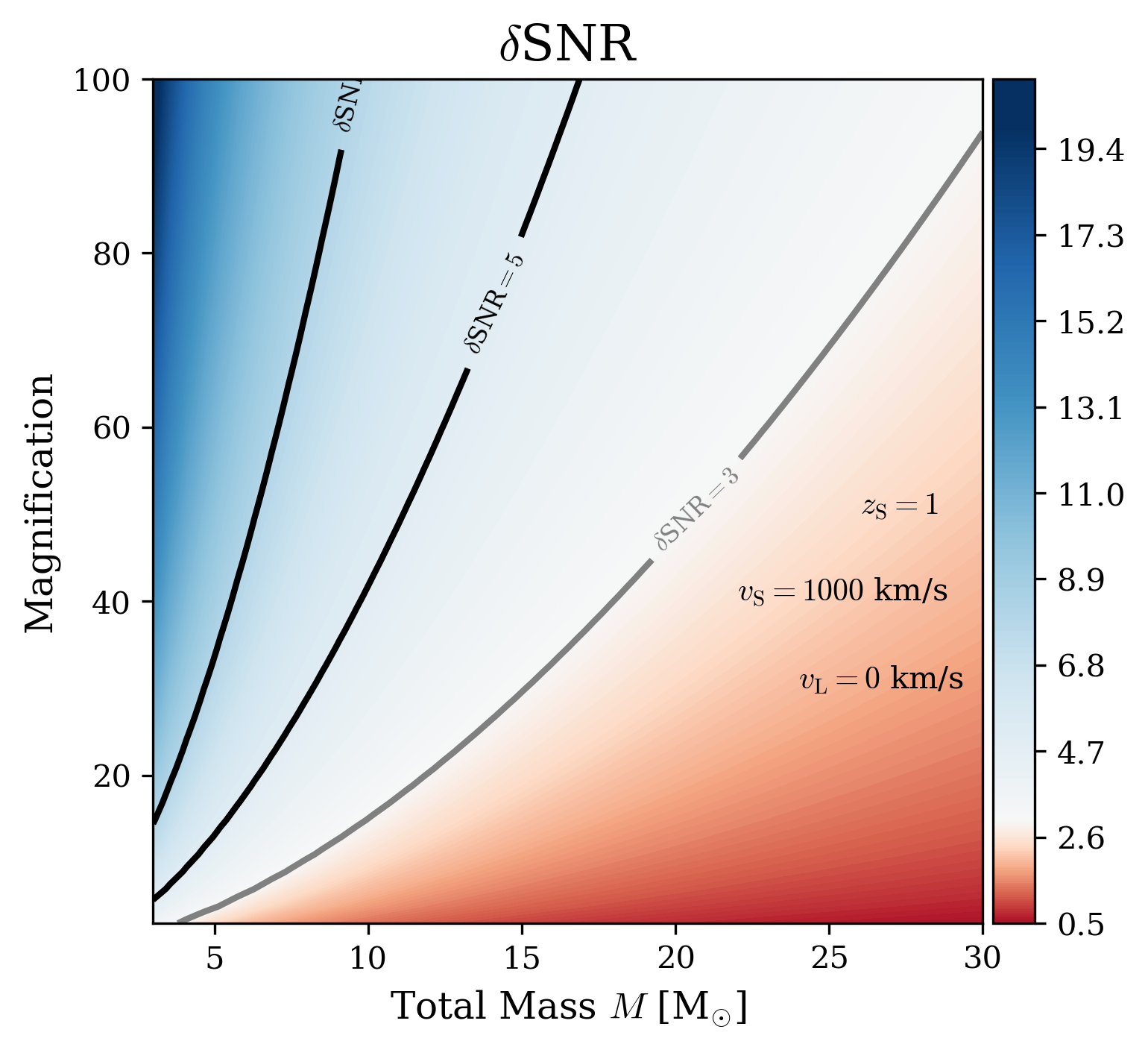}
    \includegraphics[width=0.49\textwidth]{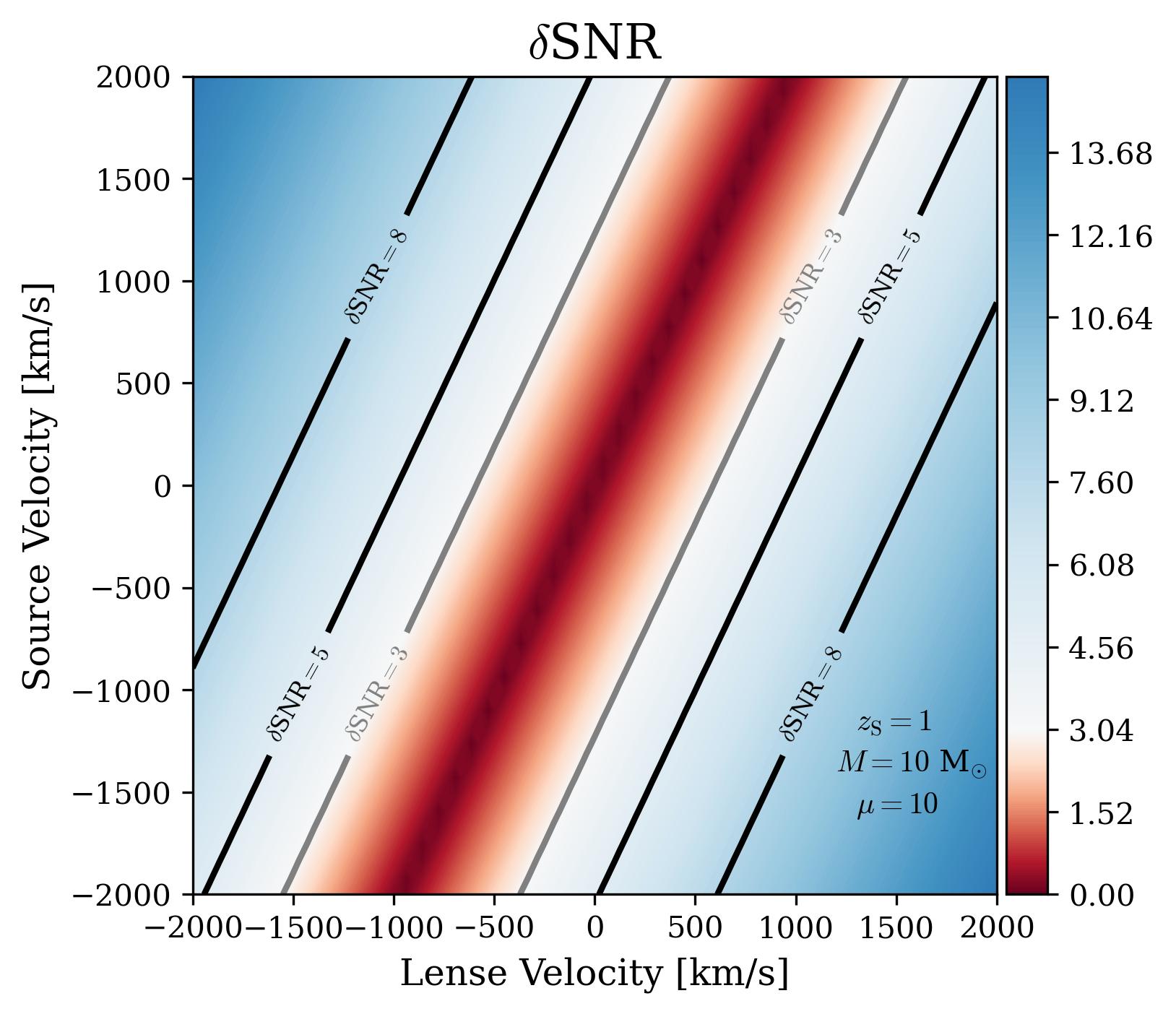}
    \caption{{\bf Einstein Telescope Detectability.} The four plots each show $\delta$SNR for measuring the relative transverse velocity induced
    GW phase shift $\delta{\phi}$, between two images of a strongly lensed GW source (Sec. \ref{sec:Einstein Telescope}).
    This GW phase shift and its detectability depends especially on the total mass of the GW source ($M$),
    the relative velocities of the lens and source $(v_L, v_S)$, the time evolving GW frequency $(f)$, the source redshift $(z_S)$,
    image magnification $(\mu)$, and the detector sensitivity through its noise power
    spectral density (see Fig. \ref{fig:dsnr}). The four plots show results for different combinations of these parameters.
    We here assume $M=2m$, $z_L = z_S/2$ and that the magnification of each lensed image is the same. 
    }
    \label{fig:GWPH_contour}
\end{figure*}

To estimate the detectability of the relative transverse motion over a large parameter space, we employ the $\delta$SNR criterion.
The latter states that a waveform perturbation, e.g. from a Doppler shift, is detectable if the following inequality is satisfied,
\begin{align}
\label{eq:dSNRcrit}
   \delta \text{SNR} &\equiv \sqrt{\left<\Delta h,\Delta h \right>} > \mathcal{C}, \\
   \Delta h&= h_{\rm ref} - h_{\rm Dopp},
   \label{eq:Deltah}
\end{align}
where $h_{\rm ref}$ here refers to a reference waveform, and $h_{\rm Dopp}$ a Doppler shifted waveform. In our setup, the GW signal from one
of the images is first chosen to be the reference waveform, where the GW signal from the other image will be the transverse
velocity induced Doppler shifted waveform.
The inner product $\left< \cdot , \cdot \right>$ represents the noise
weighted SNR $h$,
\begin{align}
    \left< h_1 , h_2\right> = 2\int_0^{\infty} \frac{\tilde{h}_1\tilde{h}_2^{*} + \tilde{h}_1^{*}\tilde{h}_2}{S_{n}(f')}{\rm{d}}f',
\end{align}
where $\tilde{h}_i$ is the Fourier space representation of the waveform $h_i$, super script `$*$' denotes the complex conjugate, 
and $S_{n}$ is the noise power spectral density of a given detector.
The interpretation of Eq. \ref{eq:dSNRcrit} is that the SNR of the difference between
unperturbed and perturbed waveforms must reach a certain threshold in order to be detectable.
The value of $\mathcal{C}$ is often chosen to be $8$, in analogy to the typical
minimum SNR required for a confident detection in LVK. In reality, its value should depend on
employed waveforms, detector properties and the specific form of the waveform perturbation.
Additionally, satisfying Eq. \ref{eq:dSNRcrit} does not guarantee that a
perturbation can be distinguished in the full parameter estimation procedure,
due to the presence of degeneracies. Nevertheless, the $\delta$SNR criterion
suffices for the purposes of this work, as it does indeed indicate whether a
waveform perturbation is in principle large enough to be detectable.
Furthermore, it allows to efficiently survey a large volume of phase space.

\subsection{Next-Generation Ground-Based Detectors}\label{sec:Einstein Telescope}

In this work we focus on results derived for next-generation ground-based detectors, as these will see hundreds of lensed GW sources,
and be highly sensitive at low GW frequencies that greatly enhances the detection of GW phase shift in general.
We focus in particular on the prospects of an instrument with a noise power spectral density similar to that of ET,
but our main results do generally also apply to CE. The sensitivity curves of ground-based detectors and a possible lensed GW signal
are illustrated in Fig. \ref{fig:dsnr}, which shows the noise curves of ET, CE, and LIGO, together with a BBH strain signal, $h_{\rm ref}$, and
the strain difference, $\Delta{h}$, between this $h_{\rm ref}$ and a similar GW signal but Doppler shifted
in phase according to Eq. \ref{eq:phi_dopp} (see figure caption).

In the following analysis, we consider a configuration of source and lens specified by the following parameters:

\begin{align}
\label{eq:paramssss}
    \text{Ref. par.:} = &\begin{cases}
        z_{\rm S}; & \text{ Source Redshift}\\
        \mathcal{M}; & \text{Chirp mass}\\
        \mu; & \text{magnification}
    \end{cases}\\ \nonumber \\
     \text{Deph. par.:} = &\begin{cases}
        z_{\rm L}; & \text{Lens Redshift}\\
        M_{\rm L}; & \text{Lens mass}\\
        v_{\rm O}; & \text{Observer velocity}
        \\
        v_{\rm L}; & \text{Lens velocity}
        \\
        v_{\rm S}; & \text{Source velocity}
    \end{cases}
\end{align}

The waveforms are generated by the analytical Newtonian result in the stationary phase
approximation \citep[][]{1994cutlerflanagan}, modified with an magnification factor from the strong lensing,
\begin{align}
    \tilde{h}(f) = \sqrt{\mu}\frac{Q}{d(z)}\left(\frac{G\mathcal{M}}{c^3}\right)^{5/6}f^{-7/6} \exp \left[ i (- \phi) \right].
    \label{eq:tilde_hf}
\end{align}
Here $f$ is the observed GW frequency, $\mathcal{M}$ is the red-shifted chirp mass, $d(z)$ is
the luminosity distance, $Q$ is a geometric pre-factor that accounts for projections of the
GW onto a given detector, $\mu$ is the lensing magnification factor, and $\phi$ is the phase
\begin{align}\label{eq:vacuum_phase}
    \phi &= 2 \pi f t(f)  +  2\left(8\pi \mathcal{M}f \right)^{-5/3},
\end{align}
where we have neglected a constant phase offset and the integration constant $\phi_{\rm c}$,
which represents the phase at coalescence. We further assume the binary evolves with
zero eccentricity. Following this notation, the waveform that is being considered dephased compared to the other is now given by,
\begin{align}
    \phi_{\rm Dopp} = \phi_{\rm ref} + \delta{\phi}(M_{\rm L},z_{\rm L},z_{\rm S},v_{\rm L},v_{\rm S},v_{\rm O}) \,.
    \label{eq:phi_dopp}
\end{align}
where $\delta{\phi}$ is the GW phase shift described by Eq. \ref{eq:dphi_general}. The $\delta$SNR is then estimated
from $\Delta{h}$, that is calculated using Eq. \ref{eq:tilde_hf} with $\phi_{\rm ref}$ and $\phi_{\rm Dopp}$ as input.

Results for ET are shown in Fig. \ref{fig:GWPH_contour}, which shows contours of $\delta$SNR as a function of the relevant parameters in our setup.
We here assume that both GW signals have the same magnification, the lens redshift relates to the source redshift as $z_L = z_S/2$,
and that the lens mass is $M_{\rm L} = 3\times10^{14}M_{\odot}$. Regarding our chosen values of parameters in the plots, we point out that current
models and observations by LIGO hint that the single BH mass spectrum peaks
around $5 \sim 10 M_{\odot}$ \citep[e.g.][]{2019ApJ...882L..24A}, the source redshift $z_s$ peaks for strong lensing
$\sim 2$ for ET \citep[e.g.][]{2022ApJ...929....9X}, galaxy clusters have a typical velocity dispersion $\mathcal{O}(1000\ kms^{-1})$,
and the lensing kernel in lens mass peaks around $10^{14}M_{\odot}$, with larger lens masses preferred for low redshift sources \citep[e.g.][]{2022ApJ...929....9X, 2023MNRAS.520..702S}.
In addition, the lensing magnification can span 2-3 orders of magnitude \citep[e.g.][]{2024PhRvD.109b4064S}.

We first consider the {\it upper left plot} which shows $\delta$SNR, as a function of $z_S$ and total rest frame GW source mass
$M$, for $v_S = 1000\ kms^{-1}$, $v_L = 0\ kms^{-1}$, and $\mu = 50$. For these values, a robust detection for $z_S \lesssim 1$ of
transverse velocity with $\delta$SNR $\sim 8$ is possible for GW sources with total mass up to $\sim 5M_{\odot}-10M_{\odot}$ with a steep dependence on $z_S$.
If one allows for a slightly lower $\delta$SNR thresshold, the available space significantly opens up, implying that GW sources up to
$z_S\sim 1$ with a total mass up in the range of $\sim 10M_{\odot}-30M_{\odot}$ will show detectable features arising from transverse velocities.
The {\it upper right plot} shows $\delta$SNR, as a function of $z_S$ and magnification $\mu$, for a fixed total GW source mass of $M=10M_{\odot}$.
As seen, for a $\delta$SNR of $\sim 5$, a magnification of $\mu \sim 40$ is required for claiming detection at $z_S\sim1$,
where $\mu \sim 100$ will bring the horizon out to $z_S \sim 2.0$.
The {\it lower left plot} shows instead the dependence on mass $M$ for fixed $z_S = 1$. This especially shows that the $\mu$
has to be greater than $\sim 40$ for systems with $M \sim 10M_{\odot}$ for our chosen relative velocity to be resolved.
The {\it lower right plot} shows the dependence on $v_L$ and $v_S$, which enter into the GW phase shift with different weight factors
set by the geometry and cosmology as seen in Eq. \ref{eq:vd}. This is clearly seen in the $\delta$SNR contours,
e.g. for $v_L = 0\ kms^{-1}$ and $v_S \pm 1000\ kms^{-1}$ the maximum $\delta$SNR is $< 3$, whereas for
$v_s = 0\ kms^{-1}$ and $v_L \pm 1000\ kms^{-1}$ the maximum $\delta$SNR is $> 5$. The two velocities can naturally also
add up constructively, resulting for our chosen setup in a $\delta$SNR $\sim 8$ if the source and the lens
move opposite to each other with a velocity of $\sim 1000\ kms^{-1}$ each, which is certainly not unreasonably.

To conclude, our considered examples definitly hint that it should be possible
to put constrains on the relative transverse velocity of GW sources that falls into the regime of real astrophysical systems.
We have only considered a small part of the relevant phase space here, but our results serve as a major
motivation to study this in greater detail.

\section{Conclusions}\label{sec:Conclusions}

We have in this paper illustrated the possibility of measuring the relative transverse motion of GW
sources in strong lens systems, when two or more images of the GW source is observed (see Fig. \ref{fig:ill_lensingBBH}).
This is possible as the strong lensed images essentially allow the observer to see the GW source from different 
lines-of-sight, each of which will show different projections of the relative velocity of the GW source. This gives rise to a differential Doppler shift
between the images, which can be translated to an observable GW phase shift \citep[see also][]{2009PhRvD..80d4009I}.

If more than two images of are observed, one can triangulate using image pairs, or a joint fit, for the transverse velocity vector
in the source plane; we therefore refer to this technique as {\it Doppler-Triangulation}.
This method provides oppotunities of getting a rare glimpse into how GW sources move relative to the cosmic flow,
which otherwise is impossible to measure using single GW signals alone. We have focused on cosmological lensed configurations,
as we know that strongly lensed GW sources will be observed in the near
future \citep[][]{2022ApJ...929....9X, 2023MNRAS.520..702S}, but the effect we are describing could happen for any other
system where lens and source move relative to each other \citep[see also][]{2024arXiv241016378Y, 2024PhRvD.109b4064S}.

Using a $\delta$SNR criterion we quantified the observable possibilities for ET, as this is expected to detect hundreds of
lensed GW systems \citep[e.g.][]{2022ApJ...929....9X, 2023MNRAS.520..702S}.
For reasonable parameters of total BBH mass $M \sim \mathcal{O}(10M_{\odot})$, relative velocity $v \sim \mathcal{O}(1000\ kms^{-1})$,
lensing magnification $\mu \sim \mathcal{O}(10)$, source redshift $z_s \sim \mathcal{O}(1)$, and lens mass $\mathcal{O}(10^{14}M_{\odot})$,
we find that measuring the resultant GW phase shift between pairwise images should definitely be possible.
This opens up tremendous opportunities for ET as well as CE to probe the relative transverse flow of GW sources,
which undoubtedly will give better hints to where and how these systems are forming in our Universe.

It is natural to follow up on our study with a more detailed analysis using a realistic lens model, lens and source distributions,
as well as including more than two images to quantify how well one can triangulate for the full relative transverse velocity vector.
It is also of high interest to quantify, possibly using state-of-the-art cosmological simulations, if different GW source populations are
expected to have statistically different velocity distributions as a function of redshift.
Using simulations and observational constraints, combined with GW source population codes,
we plan on looking into this in upcoming papers.

\section{Acknowledgments}

The authors are grateful Juan Urrutia, Mikołaj Korzyński, Miguel Zumalacárregui, and Graham Smith, for useful discussions.
We further thank the The Erwin Schrödinger International Institute for Mathematics and Physics (ESI)
and the organizers of the workshop "Lensing and Wave Optics in Strong Gravity" where part of this work was
carried out.
K.H, L.Z., P.S., and J.S. are supported by the Villum Fonden grant No. 29466, and by the ERC Starting
Grant no. 101043143 -- BlackHoleMergs led by J. Samsing.
J.M.E, R.L and L.V. are supported by the research grant no. VIL37766 and no. VIL53101 from Villum Fonden, and the DNRF Chair program grant no. DNRF162 by the Danish National Research Foundation. 
J.M.E. is also supported by the Marie Sklodowska-Curie grant agreement No.~847523 INTERACTIONS. 
The Tycho supercomputer hosted at the SCIENCE HPC center at the University of Copenhagen was used for supporting this work.

\bibliographystyle{aasjournal}
\bibliography{NbodyTides_papers}

\end{document}